%
%
%
%
%
%
%
%
%
%
%
%
\hoffset=0.0in
\voffset=0.0in
\hsize=6.5in
\vsize=8.9in
\normalbaselineskip=12pt
\normalbaselines
\topskip=\baselineskip
\parindent=15pt
\parskip=0pt plus 1.2pt
%
%
%
\let\al=\alpha
\let\bt=\beta

\let\dl=\delta
\let\ep=\epsilon

\let\la=\langle
\let\ra=\rangle
\let\pa=\partial
\let\lf=\left
\let\rt=\right
\let\dt=\cdot
\let\del=\nabla
\let\dg=\dagger
\let\ddg=\ddagger
\let\q=\widehat

\let\h=\hbar

\let\rta=\rightarrow

\let\x=\times
\let\dy=\displaystyle

\let\sy=\scriptstyle

\let\:=\>
\let\\=\cr
\let\emph=\e

\let\m=\hbox

\let\cl=\centerline

\def\e#1{{\it #1\/}}
\def\textbf#1{{\bf #1}}
\def\[{$$}
\def\]{\[}
\def\re#1#2{$$\matrix{#1\cr}\eqno{(#2)}$$}
\def\de#1{$$\matrix{#1\cr}$$}

\def\eqdf{\buildrel{\rm def}\over =}
\def\hf{{\sy {1 \over 2}}}
\def\qr{{\sy {1 \over 4}}}

\def\qH{\q{H}}

\def\mathrm#1{{\rm #1}}

\def\mathcal#1{{\cal #1}}

\def\mbf{\fam\bffam\tenbf}
\def\bv#1{{\mbf #1}}

\def\vr{\bv{r}}

\def\vp{\bv{p}}

\def\vE{\bv{E}}
\def\vB{\bv{B}}
\def\vA{\bv{A}}
\def\vF{\bv{F}}
\def\vG{\bv{G}}

\def\vj{\bv{j}}
\def\vk{\bv{k}}

\def\vz{\bv{0}}
\def\vPhi{\bv{\Phi}}
\def\vPi{\bv{\Pi}}
\def\vPsi{\bv{\Psi}}

\def\qvp{\q{\vp}}
\def\qvA{\q{\vA}}

\def\qvPsi{\q{\vPsi}}

\font\frtbf = cmbx12 scaled \magstep1
\font\twlbf = cmbx12
\font\ninbf = cmbx9
\font\svtrm = cmr17
\font\twlrm = cmr12
\font\ninrm = cmr9

\def\abstract#1{{\ninbf\cl{Abstract}}\medskip
\openup -0.1\baselineskip
{\ninrm\leftskip=2pc\rightskip=2pc #1\par}
\normalbaselines}
\def\sct#1{\vskip 1.33\baselineskip\noindent{\twlbf #1}\medskip}

\def\so{\raise 0.65ex \m{\sevenrm 1}}
\def\sk{\par\vskip 0.66\baselineskip}
{\svtrm
\cl{Source-free electromagnetism's canonical fields}
\medskip
\cl{reveal the free-photon Schr\"{o}dinger equation}
}
\bigskip
{\twlrm
\cl{Steven Kenneth Kauffmann}
}
\cl{American Physical Society Senior Life Member}
\medskip
\cl{43 Bedok Road}
\cl{\#01-11}
\cl{Country Park Condominium}
\cl{Singapore 469564}
\cl{Tel \&\ FAX: +65 6243 6334}
\cl{Handphone: +65 9370 6583}
\smallskip
\cl{and}
\smallskip
\cl{Unit 802, Reflection on the Sea}
\cl{120 Marine Parade}
\cl{Coolangatta QLD 4225}
\cl{Australia}
\cl{Tel/FAX: +61 7 5536 7235}
\cl{Mobile:  +61 4 0567 9058}
\smallskip
\cl{Email: SKKauffmann@gmail.com}
\bigskip\smallskip
\abstract{
Classical equations of motion that are first-order in time and conserve energy
can only be quantized after their variables have been transformed to canonical
ones, i.e., variables in which the energy is the system's Hamiltonian.  The
source-free version of Maxwell's equations is purely dynamical, first-order in
time and has a well-defined nonnegative conserved field energy, but is
decidedly noncanonical.  That should long ago have made source-free Maxwell
equation canonical Hamiltonization a research priority, and afterward, standard
textbook fare, but textbooks seem unaware of the issue.  The opposite parities
of the electric and magnetic fields and consequent curl operations that typify
Maxwell's equations are especially at odds with their being canonical fields.
Transformation of the magnetic field into the transverse part of the vector
potential helps but is not sufficient; further simple nonnegative symmetric
integral transforms, which commute with all differential operators, are needed
for both fields; such transforms also supplant the curls in the equations of
motion.  The canonical replacements of the source-free electromagnetic fields
remain transverse-vector fields, but are more diffuse than their predecessors,
albeit less diffuse than the transverse vector potential.  Combined as the real
and imaginary parts of a complex field, the canonical fields prove to be the
transverse-vector wave function of a time-dependent Schr\"{o}dinger equation
whose Hamiltonian operator is the quantization of the free photon's square-root
relativistic energy.  Thus proper quantization of the source-free Maxwell
equations is identical to second quantization of free photons that have normal
square-root energy.  There is no physical reason why first and second
quantization of any relativistic free particle ought not to proceed in precise
parallel, utilizing the square-root Hamiltonian operator.  This natural
procedure leaves no role for the completely artificial Klein-Gordon and Dirac
equations, as accords with their grossly unphysical properties.
}

\sct{Introduction}
\noindent
Notwithstanding the approximately century and a half which has passed
since both the development of Hamiltonian classical dynamics and also
the final codification of the laws which govern electromagnetic fields
in configuration space, canonical Hamiltonian formulation of the pure%
ly dynamical source-free instance of electromagnetic field theory in
configuration space unaccountably still lingers as essentially terra
incognita.  This is all the stranger insofar as proper canonical Ham%
iltonization of a classical dynamical system is a fundamental prereq%
uisite to its rigorous unambiguous quantization, whether by the Hamil%
tonian phase-space path integral~[1], or by the self-consistent
(slight) extension of Dirac's canonical commutation rule~[2].

A crucial aspect of correct canonical Hamiltonization of the source-%
free instance of Maxwell's field equations in configuration space is
that it is subject \e{not only} to those equations \e{themselves}, but
\e{also} to the fact that the conserved energy content of the source-%
free (transverse) dynamical electromagnetic fields is a very speci%
fic \e{known functional} of those fields.  This specific conserved
field-energy functional \e{necessarily} becomes the field system's
\e{Hamiltonian functional} when its fields have been transformed to
\e{canonical} ones, i.e., we shall have transformed the fields to can%
onical ones when the field equations expressed in terms of the trans%
formed fields \e{exactly match} the Hamiltonian equations of motion
for the transformed fields which result from taking the Hamiltonian
to be the conserved field-energy functional as expressed in terms of
the transformed fields.  Therefore the fact that the conserved field-%
energy functional is \e{known a priori} places a \e{strong constraint}
on just what the canonical fields can be.

In the source-free case, Maxwell's field equations in configuration
space are strictly linear and homogeneous, so it is not surprising
that we can restrict ourselves to purely linear transformations of the
dynamical magnetic and transverse electric fields in the search for
properly canonical fields.  (The longitudinal electric field is \e{%
never} dynamical in character, and it \e{vanishes identically} in the
source-free case.)  It as well turns out to be unnecessary to mix the
magnetic field with the transverse electric field in the course of
this search.  But the axial/polar dichotomy between the magnetic field
and the transverse electric field turns out to be incompatible with
fields that are properly canonical; this failing is readily rectified
by a one-to-one orthogonal linear mapping of the axial-vector magnetic
field onto a transverse polar-vector counterpart.  The resulting
transformed field equations, however, still fail to match the Hamil%
tonian equations of motion which follow from the transformed conserved
field-energy functional, but a further nonnegative symmetric linear
transformation of all fields, which commutes with all partial deriva%
tive operators and doesn't affect the field equations, repairs this
disparity.  This last transformation is decidedly \e{nonlocal} howev%
er: the still transverse \e{canonical} fields turn out to be \e{dif%
fused} relative to the transverse electric and magnetic fields---albe%
it to a \e{lesser extent} than the transverse \e{vector potential} is
\e{likewise diffused} relative to the magnetic field (the Aharanov-%
Bohm effect is a quite dramatic manifestation of that particular rela%
tive diffusion).

The transverse canonical vector fields are not only diffused relative
to the transverse dynamical electric and magnetic fields; they also
have a different dimensionality, namely that of the square root of ac%
tion density rather than that of the square root of energy density,
which is the dimensionality of electric and magnetic fields.  If the
two transverse canonical vector fields are divided by $(2\h)^\hf$ and
then combined as the real and imaginary parts of a complex transverse%
-vector field, it is seen that this complex-valued field satisfies the
natural time-dependent Schr\"{o}dinger equation in configuration rep%
resentation whose Hamiltonian operator is $\h c(-\del^2)^\hf$, namely
$|c\qvp|$, the Hamiltonian operator for a first-quantized free soli%
tary ultrarelativistic zero-mass particle that is the zero-mass limit
of the natural correspondence-principle-mandated square-root relati%
vistic free-particle Hamiltonian $(m^2c^4 + |c\qvp|^2)^\hf$.  Moreov%
er, the dimensionality of the complex-valued transverse-vector field
in this Schr\"{o}dinger equation is that of the square root of proba%
bility density, which is appropriate for a Schr\"{o}dinger equation
wave function.  In other words, the two transverse \e{canonical} vec%
tor fields of source-free electromagnetic field theory comprise the
\e{natural basis} for the complex-valued transverse-vector \e{wave-%
function description} of the \e{first-quantized free solitary photon}.
Source-free electromagnetic field theory \e{is} in this way precisely
the time-dependent Schr\"{o}dinger-equation description of the first-%
quantized free solitary photon with its natural square-root Hamilton%
ian operator, but that stark fact simply \e{does not become technical%
ly manifest} until this field theory has been \e{properly canonically
Hamiltonized}!

It is delightfully amazing that James Clerk Maxwell, building atop the
foundation laid by Michael Faraday's experimental results, effectively
discovered the Schr\"{o}dinger equation, indeed for a tricky trans%
versely-polarized spin 1 ultrarelativistic particle, very long before
Erwin Schr\"{o}dinger's own nonrelativistic quantum insights.  Remark%
ably, Maxwell could accomplish this feat with no knowledge whatsoever
of the quantum of action---whose discovery by Max Planck still lay
well in the future---because of the intriguing happenstance that in
configuration representation the photon's zero-mass property ``releas%
es'' a factor of $\h$ from its Hamiltonian operator that neatly
\e{cancels out} the factor of $\h$ which is a \e{fixture} of the left%
-hand side of the time-dependent Schr\"{o}dinger equation.  Thus does
ultrarelativistic quantum mechanics chameleon-like metamorphose into
``classical field theory''!  The \e{technical details} of this connec%
tion \e{can not}, of course, be fully laid bare until Maxwell's effec%
tive formulation of the theory in terms of the axial magnetic and
transverse-polar electric fields has been \e{properly canonically Ham%
iltonized}---the two corresponding transverse canonical fields have
the same (not the opposite) parity, are relatively somewhat more dif%
fused than their antecedent magnetic and electric fields, albeit less
so than is the case for the transverse vector potential relative to
its antecedent magnetic field, and are joined together in a single
complex-valued transverse-vector wave function that describes the
first-quantized free solitary photon.

In the world of quantum theory, canonical Hamiltonization becomes
merely the required \e{prelude} to quantization; this of course also
applies to the quantization of source-free electromagnetic field theo%
ry.  After completion of canonical Hamiltonization, canonical quanti%
zation parlays the Poisson bracket relations of the fundamental canon%
ical dynamical variables into commutation relations.  In canonically
Hamiltonized source-free electromagnetic field theory, the fundamental
canonical dynamical variables are the vector components of the two
transverse-vector canonical fields themselves.  Thus it is the compon%
ents of these transverse-vector canonical fields themselves that quan%
tization promotes into noncommuting Hermitian operators; these opera%
tors inherit from their simple mutual Poisson bracket relations equal%
ly simple mutual commutation relations.  The vector components of the
free solitary-photon wave functions, being straightforward complex
linear combinations of the corresponding vector components of the two
real transverse-vector canonical fields, become non-Hermitian noncom%
muting operators such that there are as well simple commutation rela%
tions between the vector components of the quantized wave function and
those of its \e{Hermitian conjugate}(the \e{transverse character} of
these canonical vector fields and complex vector wave functions sub%
tracts an annoying longitudinal projection-operator term from their
formal configuration-representation commutators, which creates an ugly
and potentially confusing distraction that is merely technical in na%
ture).  The Hamiltonian functional \e{also} becomes quantized via its
bilinear dependence on the real canonical fields (or, as well, via its
linear dependence on both the complex photon wave function and its
complex conjugate), and is a Hermitian operator.  As is the normal
case in quantum theory, the Heisenberg picture with respect to this
Hamiltonian operator manifests \e{the same dynamical equations for
the quantized field operators as they obeyed prior to their quantiza%
tion}.  Thus the \e{quantized} wave function operator \e{still satis%
fies the very same Schr\"{o}dinger equation} that it satisfied \e{pri%
or} to its quantization, \e{which can thus properly be termed second
quantization}.  This Schr\"{o}dinger equation in its \e{second-quan%
tized} form obviously \e{still features the very same} $m\rta 0$ limit
of the relativistic \e{square-root} Hamiltonian operator $(m^2c^4 + |c
\qvp|^2)^\hf$ (which comes to $\h c(-\del^2)^\hf$ in configuration
representation) that it featured \e{prior} to that quantization!  Fin%
ally, the simple commutation relation between the quantized non-Her%
mitian wave function operator and its Hermitian conjugate is such that
the quantized wave function operator is interpretable as the annihila%
tion operator for free photon states in the underlying Hilbert space
(commonly called Fock space), while its Hermitian conjugate is like%
wise interpretable as the creation operator for such free photon
states.

This \e{second-quantized} theory of \e{arbitrarily many} free photons,
or, \e{equivalently}, quantized theory of source-free electromagne%
tism, is not, of course, the end of the physics story.  We have so far
\e{postponed} consideration of the coupling of \e{charged matter} to
the dynamical transverse part of the electromagnetic fields, i.e., to
photons.  In the rather global technical language of the Maxwell equa%
tions, this coupling (in fact an inhomogeneous driving term) can be
mathematically abstracted in terms of the transverse part of the cur%
rent-density vector field.  From a microscopic charged-particle pers%
pective, however, relativistic coupling to electromagnetism occurs via
the four-vector potential $A^\mu$.  The part of $A^\mu$ which is re%
lated to the dynamical transverse electromagnetic fields is, naturally
enough, the transverse part of the vector potential $\vA$, which we
denote as $\vA_T$.  Indeed the precise linear relationship of $\vA_T$
to the complex-valued transverse-vector photon wave function and its
complex conjugate can be readily traced.  The \e{remaining} $A^0$ and
\e{longitudinal} part of $\vA$ are related 1) to the nondynamical
longitudinal part of the electric field, which is a mere homogeneous
functional of the global charge density, and 2) to a nonphysical
``gauge degree of freedom''.  Thus they bear no direct relation to the
photon wave function.  Because the photon wave function and its com%
plex conjugate become operators upon second quantization, $\vA_T$ be%
comes a linearly related operator as well, in fact a Hermitian one
that can either create or annihilate a free photon state in the under%
lying Fock space.  This detailed tying of $\vA_T$ to the free-photon
annihilation and creation operators is the technical way in which the
direct interaction between photons and charged particles finds effect
in interacting second-quantized theory.  Such a theory must of course
as well includes the above-described second-quantized Hamiltonian
functional operator of source-free electromagnetism, i.e., the Hamil%
tonian operator for the for an arbitrary number of noninteracting free
photons.

What about $A^0$ and the longitudinal part of $\vA$, which we write as
$\vA_L$?  These are also coupled to charged particles, but have no di%
rect relation to \e{dynamical} transverse electromagnetism or photons.
They encompass the nondynamical physics of $\vE_L$, the longitudinal
part of the electric field, which can be deduced from Maxwell's equa%
tions to be \e{entirely the creature of the global charge density},
and therefore is properly termed \e{coulombic}, and the ``nonphysics''
of the choice of gauge function.  Given the global Hamiltonian func%
tional for the charged particles, specifically \e{including in partic%
ular} their formal interaction with electromagnetism via arbitrary
$A^\mu$, we readily obtain the global charge density $\rho$ as the
functional derivative of that Hamiltonian functional with respect to
$A^0$.  The longitudinal part of the electric field, namely $\vE_L$,
is then obtained as a closed-form linear homogeneous functional of
$\rho$, but this is not sufficient to pin down \e{both} $A^0$ and
$\vA_L$, which, additionally, calls for a choice of gauge.  The sim%
plest choice is the Coulomb gauge, which rather brusquely puts $\vA_L$
to zero, oblivious to even the slightest pretense of special-relati%
vistic finesse.  Thereupon $A^0$ also becomes a closed-form linear
homogeneous functional of $\rho$ via the static Coulomb kernel.  Aside
from the blatantly nonrelativistic character of a potential $A^0$ that
arises from the static Coulomb kernel, two other subtleties present
themselves: 1) since $A^0$ couples to the very same particles that
give rise to the global $\rho$, one must compensate for double count%
ing by the usual expedient of halving $A^0$ relative to its naive val%
ue, and 2) if it should turn out that $\rho$ itself still retains a
formal dependence on $A^0$, one will have only obtained an \e{implicit
equation} for $A^0$, one that requires solution (probably only an ap%
proximate one via successive iteration will be feasible).

A relativistically more plausible gauge~[3] is rooted in the Lorentz
condition $(c\del\dt\vA_L + \dot A^0) = 0$ conjoined to the require%
ment that $A^0$ be linear and homogeneous in $\rho$, but \e{retarded}
by $(|\vr-\vr'|/c)$  This can be achieved because the Lorentz condi%
tion imposes on $A^0$ the requirement that it be related to $\rho$ by
the $c$-speed compatible second-order partial differential equation
$(\ddot A^0/c^2 - \del^2A^0) = \rho$.  The upshot is that $A^0$
is a closed-form integral that is linear and homogeneous in $\rho$
via the static Coulomb kernel, but for $\rho$ evaluated at the \e{re%
tarded time} $(t - |\vr -\vr'|/c)$.  With $A^0$ in hand, $\vA_L$ can
be determined from $\dot A^0$ via the Lorentz condition.  In the spe%
cial case that $\dot\rho = 0$, this ``retarded Lorentz gauge'' is
identical to the Coulomb gauge.  The above caveats concerning halving
$A^0$ relative to its naive value and the possibility of obtaining
only an implicit equation for $A^0$ still apply.

The following sections provide the technical/mathematical details that
explicitly underpin and demonstrate what is described and discussed in
the foregoing paragraphs.

\sct{Longitudinal nondynamical and transverse dynamical electomagnetic fields}
\noindent
The four Maxwell equations for the electromagnetic field $(\vE, \vB)$
with four-current source $(\rho, \vj/c)$ are comprised of Coulomb's law,
\re{\del\dt\vE = \rho,
}{1a}
Faraday's law,
\re{\del\x\vE = -\dot{\vB}/c,
}{1b}
Gauss' law,
\re{\del\dt\vB = 0,
}{1c}
and Maxwell's law,
\re{\del\x\vB = (\vj + \dot{\vE})/c.
}{1d}

The Coulomb and Gauss laws do not involve time derivatives of the el%
ectromagnetic fields, i.e., they are nondynamical in character.  This
raises the possibility that some part of the electromagnetic field
$(\vE, \vB)$ may itself be of nondynamical character, i.e. determinab%
le without reference to any initial conditions.  This in fact turns
out to be the case for the longitudinal part of the electric field,
which it is therefore very useful to detach from the rest of the elec%
tromagnetic field.  The ability to separate a vector field into its
longitudinal and transverse parts in a linear and unique fashion is
in fact extremely useful \e{throughout} electromagnetic theory, so we
turn first to a discussion of how that is carried out.

We shall now indicate that any vector field $\vF(\vr)$ which is con%
tinuously differentiable, and for which $|\del\dt\vF(\vr)|/|\vr|$ is
intgrable over all space, has a unique decomposition \e{achieved by
linear operation} into the \e{sum} of its longitudinal part
$\vF_L(\vr)$ with its transverse part $\vF_T(\vr)$, where
$\del\dt\vF_L = \del\dt\vF$, $\del\x\vF_L = \vz$ and
$\del\dt\vF_T = 0$.

Since we require that $\del\x\vF_L = \vz$, it will need to be the case
that $\vF_L(\vr) = -\del S(\vr)$, where $S(\vr)$ is a scalar field.
Since we also require that $\del\dt\vF_L = \del\dt\vF$, we must have
that $-\del^2 S = \del\dt\vF$, whose general solution is,
\de{
S(\vr) = c_0 + \vk_0\dt\vr + [(-\del^2)^{-1}(\del\dt\vF)](\vr),
}
where $c_0$ and $\vk_0$ are arbitrary constants, and the natural nota%
tion $(-\del^2)^{-1}$ denotes the integral operator whose configura%
tion-representation kernel $\la\vr|(-\del^2)^{-1}|\vr'\ra$ is,
\de{
\la\vr|(-\del^2)^{-1}|\vr'\ra = (4\pi|\vr - \vr'|)^{-1},
}
because, as is well-known,
\de{
(-\del^2_\vr)(4\pi|\vr - \vr'|)^{-1} = \dl^{(3)}\!(\vr -\vr') = \la\vr|\vr'\ra.
}
Since $\vF_L = -\del S$, we obtain $\vF_L = \vk_0 -
\del[(-\del^2)^{-1}(\del\dt\vF)]$, where $\vk_0$ is an arbitrary con%
stant.  Because we require the decomposition of $\vF$ into the sum of
$\vF_L$ with $\vF_T$ to be achieved by \e{linear operation}, $\vk_0$
must vanish identically, and therefore,
\de{
\vF_L = -\del[(-\del^2)^{-1}(\del\dt\vF)],
}
or
\de{
\vF_L(\vr) =
  -\del_\vr\int(4\pi|\vr-\vr'|)^{-1}(\del_{\vr'}\dt\vF(\vr'))\,d^3\vr'.
}
We now note from the above two equations the \e{key fact} that $\vF_L$
is \e{entirely determined} by $\del\dt\vF$.  Since, of course, $\vF_T
= \vF - \vF_L$, the fact that $\del\dt\vF_L = \del\dt\vF$ implies that
$\del\dt\vF_T = 0$, as is required for $\vF_T$.

We shall now systematically apply this \e{linear decomposition} into
its longitudinal and transverse parts to \e{each vector field} in each
of the four Maxwell equations of Eqs.~(1).  We consequently obtain,
\re{
\del\dt\vE_L = \rho,
}{2a}
\re{
\del\x\vE_T = -\dot\vB_T/c,
}{2b}
\re{
\dot\vB_L = \vz,
}{2c}
\re{
\vB_L = \vz,
}{2d}
\re{
\del\x\vB_T = (\vj_T + \dot\vE_T)/c,
}{2e}
and,
\re{
\vj_L + \dot\vE_L = \vz.
}{2f}
From Eq.~(2a) and the representation for the longitudinal part of any
vector field obtained above, we obtain $\vE_L$ in closed form,
\re{
\vE_L(\vr, t) = -\del_\vr[(-\del^2)^{-1}\rho](\vr) =
             -\del_\vr\int(4\pi|\vr-\vr'|)^{-1}\rho(\vr', t)\,d^3\vr'.
}{3a}
Eq.~(2d) makes Eq.~(2c) redundant.  In consequence of Eq.~(2d), $\vB =
\vB_T$, which we shall simply \e{always bear in mind}.  That enables
us to drop all references to $\vB_T$.  Therefore Eq.~(2b) can be writ%
ten,
\re{
\del\x\vE_T = -\dot\vB/c,
}{3b}
and Eq.~(2e) can likewise be written,
\re{
\del\x\vB = (\vj_T + \dot\vE_T)/c.
}{3c}
With regard to Eq.~(2f), we recall that the longitudinal part of a
vector field is \e{completely determined} by its divergence.  There%
fore it is sufficient to simply work with the divergence of Eq.~(2f),
from which we deduce that $\del\dt\vj + \del\dt\dot\vE_L$ = 0.  Now
there is \e{nothing more} to be learned about $\vE_L$, as it is given
\e{in closed form in terms of} $\rho$ in Eq.~(3a).  Therefore we use
Eq.~(2a) to eliminate its presence, and thus obtain the celebrated
\e{constraint} due to charge/current conservation:
\re{
\del\dt\vj + \dot\rho = 0.
}{3d}
Eqs.~(3), plus the fact that $\vB$ is purely transverse, completely
replace the Maxwell equation system of Eqs.~(1), and are clearly far
more informative than that system.  In particular, it is crystal-%
clear from Eq.~(3a) that $\vE_L$ is a completely nondynamical vari%
able which is \e{entirely independent of the choice of initial condi%
tions}.  It is \e{only the transverse fields} $\vE_T$ and $\vB$ that
are actual dynamical variables, and \e{only these} can legitimately
be incorporated into a standard dynamical framework!

\sct{Source-free electromagnetic theory}
\noindent
We shall now do away with the charged matter sources, namely put
$\rho$ and $\vj$ to zero, and thereby deal with purely self-sustain%
ing radiation.  Eq.~(3a) thereupon becomes $\vE_L = \vz$, and thus can
be dropped entirely along with Eq.~(3d), which reduces to the trivial
identity $0 = 0$. We are left with the following two dynamical equa%
tions, which involve only the two purely transverse fields $\vE_T$ and
$\vB$,
\re{
\dot\vB = -c\del\x\vE_T,
}{4a}
and,
\re{
\dot\vE_T = c\del\x\vB.
}{4b}
In addition to these equations of motion, source-free electromagne%
tism has a very well-known conserved nonnegative field-energy func%
tional~[4,5], which is given by,
\re{
E[\vE_T, \vB] = \hf\int\lf[|\vE_T(\vr, t)|^2 + |\vB(\vr, t)|^2\rt]\,d^3\vr.
}{4c}
We can readily calculate the time rate of change of $E[\vE_T, \vB]$
by applying the two field equations of motion, namely Eqs.~(4a) and
(4b), which yields,
\re{
dE/dt = c\int\lf[(\del\x\vB)\dt\vE_T - \vB\dt(\del\x\vE_T)\rt]\,d^3\vr
}{4d}
Now it is an \e{identity} that,
\de{
\del\dt(\vB\x\vE_T) = (\del\x\vB)\dt\vE_T - \vB\dt(\del\x\vE_T),
}
which implies that,
\re{
dE/dt = c\int\del\dt(\vB\x\vE_T)\,d^3\vr,
}{4e}
and the integral over all space of such a divergence will, of course,
vanish under normal circumstances.  Thus the nonnegative field-energy
functional $E[\vE_T, \vB]$ is indeed conserved,
\re{
dE/dt = 0.
}{4f}
A useful corollary of this demonstration is that under normal circum%
stances,
\re{
\int(\del\x\vF)\dt\vG\>d^3\vr = \int\vF\dt(\del\x\vG)\,d^3\vr.
}{5}
Now the correct conserved energy of a dynamical system becomes the
system's Hamiltonian whenever that system is described by correct can%
onical variables.  Here it is immediately clear that $(\vE_T, \vB)$
are \e{not} correct canonical fields for this dynamical electromagne%
tic system, because treating the system's correct energy functional of
Eq.~(4c) as the system's Hamiltonian functional results in the puta%
tive Hamiltonian equations of motion $\dot\vE_L = \vB$ and $\dot\vB =
-\vE_L$, which blatantly disagree with the actual equations of motion
that are given by Eqs.~(4b) and (4a)---there is \e{no trace} of the
very prominent curl operations of the equations of motion to be found
in the much more austerely straightforward Hamiltonian equations that
flow from the nonnegative field-energy functional.

\sct{Like parities and additional diffuseness of the canonical fields}
\noindent
We are now clearly obliged to search for correct canonical fields
for this source-free electromagnetic system by trying out transforma%
tions of $(\vE_T, \vB)$.  Since we are dealing with linear equations
and a bilinear energy functional, it is clear that we can restrict
ourselves to linear transformations.  Both the form of the bilinear
energy functional and that of the present equations of motion of
Eqs.~(4b) and (4a) strongly suggest that we not look at any transfor%
mations that mix the electric with the magnetic field.  In light of
the form of the bilinear energy functional, however, it seems urgent
to find a transformation that eliminates the curl operations from
the equations of motion.  The presence of the curl operations is en%
twined with the fact that $\vB$ is an axial transverse vector while
$\vE_L$ is a polar transverse vector; thus it seems important to find
a way map $\vB$ onto a polar transverse vector without harming any
physically important information carried by $\vB$.

The vector potential $\vA$ is a polar vector whose tranverse part
appears to carry all the physical information that is contained in
the transverse axial vector $\vB$.  This is so because $\vB =
\del\x\vA = \del\x\vA_T$.  However, $\vA_T$ itself has a different
dimensionality from $\vB$, and is also considerably more diffuse, as
the Aharanov-Bohm effect pointedly illustrates.  It would be good to
be able to apply a transformation to $\vA_T$ that leaves its trans%
verse polar nature intact, yet compensates for its differences from
$\vB$ in dimensionality and diffuseness.  In this regard, it is inter%
esting to note that $\del\x\vB = \del\x(\del\x\vA_T) = -\del^2\vA_T$,
which is \e{less} diffuse than $\vB$, and errs in dimensionality rela%
tive to $\vB$ in the \e{opposite} direction from the dimensionality
error made by $\vA_T$.  These considerations strongly suggest that the
object we would \e{truly like} to have is $(-\del^2)^\hf\,\vA_T$.
Since, as we have noted just above, $\del\x\vB = -\del^2\vA_T$, we
have that, $(-\del^2)^\hf\,\vA_T = (-\del^2)^{-\hf}\,\del\x\vB$, i.e.,
we can conveniently express the entity we want \e{entirely} in terms
of $\vB$ \e{itself}, without \e{any need} to make reference to
$\vA_T$.  It is convenient to invent a shorthand notation for this de%
sired transformation of $\vB$ from an axial transverse vector to a po%
lar transverse vector that otherwise mimics $\vB$ \e{itself} just as
\e{closely as possible},
\re{
\vB^\ddg\eqdf (-\del^2)^{-\hf}\,\del\x\vB,
}{6a}
which has the marvelous property that,
\re{
(\vB^\ddg)^\ddg = \vB,
}{6b}
i.e., this linear operation is actually a \e{conjugation} when it is
\e{restricted to transverse vector fields}.  We can dub it ``polar-ax%
ial conjugation''.  To quell any lingering doubts as to the calcula%
tional ``meat'' on its somewhat abstract ``bone'', we explicitly exhi%
bit the integral-operator kernel of $(-\del^2)^{-\hf}$ in configura%
tion represention,
\re{
{\dy\la\vr|(-\del^2)^{-\hf}\,|\vr'\ra =
(2\pi)^{-3}\int|\vk|^{-1}e^{i\vk\dt(\vr-\vr')}d^3\vk =
{1\over 2\pi^2|\vr-\vr'|^2}}.
}{6c}
It is, moreover, very satisfying to note that the physically key
transverse part of the vector potential is neatly related to this
conjugate of $\vB$,
\re{
\vA_T = (-\del^2)^{-\hf}\,\vB^\ddg.
}{6d}
Finally, just as a conjugation operation \e{ought} to be, polar-%
axial conjugation is \e{orthogonal} in the natural Hilbert space
of \e{transverse vector fields}.  Thus the conserved nonnegative
field-energy functional $E[\vE_T, \vB]$ of Eq.~(4c) is \e{invariant}
under this conjugation.  In particular, from the corollary given by
Eq.~(5), the \e{symmetric nature} of the linear operator
$(-\del^2)^{-\hf}$, the fact that \e{it commutes with differential
operators} and the fact that its \e{square equals} $(-\del^2)^{-1}$,
it follows that,
\re{
\int\vB^\ddg\dt\vB^\ddg d^3\vr =
\int\vB\dt\del\x[\del\x(-\del^2)^{-1}\vB]\,d^3\vr =\int\vB\dt\vB\,d^3\vr.
}{6e}

Although the polar-axial conjugation of $\vB$ leaves the conserved
nonnegative field-energy functional $E[\vE_T, \vB]$ of Eq.~(4c)
\e{form-invariant}, it \e{does} abolish the curl operations from
the \e{equations of motion}.  Thus Eq.~(4b) can immediately be
rewritten in terms of $\vB^\ddg$ as,
\re{
\dot\vE_T = c(-\del^2)^\hf\,\vB^\ddg.
}{7a}
Writing Eq.~(4a) in terms of $\vB^\ddg$ involves slightly more work,
in that the curl operator must first be applied to both sides before
the translation in terms of $\vB^\ddg$ can be made,
\re{
\dot\vB^\ddg = -c(-\del^2)^\hf\,\vE_T.
}{7b}
The operator $(-\del^2)^\hf$ is most easily handled in Fourier trans%
form, as it is distribution-valued (has locally singular behavior) in
configuration representation. It's kernel there is given by,
\re{
{\dy\la\vr|(-\del^2)^\hf\,|\vr'\ra =
(2\pi)^{-3}\int|\vk|e^{i\vk\dt(\vr-\vr')}d^3\vk =
\lim_{\ep\rta 0}\;{(3\ep^2-|\vr-\vr'|^2)\over\pi^2(\ep^2+|\vr-\vr'|^2)^3}}.
}{7c}

Having knocked the curl operations out of the equations of motion,
Eqs.~(7a) and (7b), we are much closer to our goal of achieving canon%
ical fields, but are not quite there yet.  However, what still needs
to be done is now relatively easy to light on.  The operator factor
with dimensionality of frequency that now uniformly emerges in both
equations of motion, namely $c(-\del^2)^\hf$ must \e{also} be persuad%
ed to show its face in the transformed field-energy functional: our
polar-axial conjugation transformation changed the equations of motion
and left the field-energy functional form-invariant, but our \e{next}
transformation must do precisely the \e{opposite}. Multiplying \e{both
of} $\vE_T$ and $\vB$ by a \e{power} of the operator $[c^2(-\del^2)]$ 
doesn't change the \e{form} of Eqs.~(7a) and (7b) at all, but it caus%
es that operator to the \e{negative of twice that power} to appear in 
the field-energy functional.  To have canonical consistency between
the equations of motion and the field-energy functional, the power of
the operator $[c^2(-\del^2)]$ that must appear explicitly in the field-%
energy functional is one half.  That requires the canonical fields to
be equal to the present $\vE_T$ and $\vB^\ddg$ fields times this opera%
tor to the power of minus one quarter.

Our canonical fields are now ``in the bag''.  They are explicitly,
\re{
\vPhi = [c^2(-\del^2)]^{-\qr}\,\vE_T\;,\qquad\quad
 \vPi = [c^2(-\del^2)]^{-\qr}\,\vB^\ddg.
}{8a}
The equations of motion don't change their form from that of Eqs.~(7a)
and (7b), notwithstanding our having changed to the canonical fields
$\vPhi$ and $\vPi$, because the common operator factor of
$[c^2(-\del^2)]^{\qr}$ simply factors through and out of those equa%
tions.  Thus the equations for $\vPhi$ and $\vPi$ are,
\re{
\dot\vPhi=c(-\del^2)^\hf\,\vPi\;,\qquad\quad\dot\vPi=-c(-\del^2)^\hf\,\vPhi.
}{8b}
When the field-energy functional is written in terms of the canonical
fields $\vPhi$ and $\vPi$, it \e{does} change its form; it now becomes
a Hamiltonian that is \e{consistent} with the equations of motion of
Eq.~(8b). This Hamiltonian is,
\re{
H[\vPhi, \vPi] = \hf\int\lf[\vPhi\dt\lf(c(-\del^2)^\hf\,\vPhi\rt) +
                              \vPi\dt\lf(c(-\del^2)^\hf\,\vPi\rt)\rt]\,d^3\vr.
}{8c}
We noted in Eq.~(6d) that $\vA_T = (-\del^2)^{-\hf}\,\vB^\ddg$.  Now
from Eq.~(8a) we deduce that $\vB^\ddg = c^\hf (-\del^2)^\qr\,\vPi$.
Therefore we obtain that,
\re{
\vA_T = c^\hf (-\del^2)^{-\qr}\,\vPi
}{8d}

Eq.~(8a) shows that the canonical fields are somewhat more diffuse
than the corresponding electric and magnetic fields, and while
those fields have dimensionality of the square root of energy
density, the \e{canonical} fields have dimensionality of the square
root of \e{action density}.

\sct{The complex-valued photon wave function and Schr\"{o}dinger equation}
\noindent
Now let us make the transition to the
standard \e{complex} transverse vector field which has the dimen%
sionality of the square root of \e{probability density}.  This
field is,
\re{
\vPsi = (\vPi - i\vPhi)/(2\h)^\hf.
}{9a}
We note from Eq.~(8b) that,
$$(\dot\vPi - i\dot\vPhi) = c(-\del^2)^\hf\,(-\vPhi - i\vPi) = 
-ic(-\del^2)^\hf\,(\vPi - i\vPhi).$$
Multiplying both sides of the above result through by $i\h$,
we deduce from Eq.~(9a) that,
\re{
i\h\dot\vPsi = \h c(-\del^2)^\hf\,\vPsi.
}{9b}
Eq.~(9b) is a Schr\"{o}dinger equation with square-root Hamiltonian
operator $\h c(-\del^2)^\hf = |c\qvp|$, which is that of a massless
free particle.  This Schr\"{o}dinger equation's wave function $\vPsi$
has the dimensionality of the square root of probability density, and
it is a complex transverse vector field.  In other words, after its
proper canonical Hamiltonization and field complexification, source-%
free electromagnetism perfectly describes the solitary, first-quantized
free photon, i.e., it is revealed to be first-quantized photodynamics.
Let us now tie down the last detail of this identification by rewrit%
ing the canonical Hamiltonian functional $H[\vPhi, \vPi]$ of Eq.~(8c)
in terms of $\vPsi$ and its complex conjugate $\vPsi^\ast$.  The
result is,
\re{
H[\vPsi^\ast,\vPsi]=\int\vPsi^\ast\dt\lf(\h c(-\del^2)^\hf\,\vPsi\rt)\,d^3\vr.
}{9c}
This is indeed the Hamiltonian functional that corresponds to the
configuration-space Schr\"{o}dinger equation of Eq.~(9b), whose com%
plex-valued transverse-vector wave function $\vPsi$ is now completely
ready for second quantization.  Before carrying this out, let us ex%
press $\vA_T$, the transverse part of the vector potential, in terms
of the complex-valued photon wave function $\vPsi$.  To do this we
combine the result of Eq.~(8d) with $\vPi = (\h/2)^\hf\,(\vPsi +
\vPsi^\ast)$, which follows from Eq.~(9a), to obtain,
\re{
\vA_T =(\h c/2)^\hf(-\del^2)^{-\qr}\,(\vPsi + \vPsi^\ast).
}{9d}
This reveals the transverse part of the vector potential to be some%
what more diffuse than the photon wave function, which itself is re%
vealed by Eqs.~(9a) and (8a) to be equally more diffuse than the elec%
tromagnetic fields.

\sct{Second-quantized photodynamics}
\noindent
If the photon wave function $\vPsi(\vr)$ were a complex-valued
\e{full}-vector field, it would be quantized by being changed into a
non-Hermitian operator $\qvPsi(\vr)$ whose commutation
relation with its Hermitian conjugate $\qvPsi^\dg(\vr)$ would be given
by,
\re{
[(\qvPsi(\vr))_i, (\qvPsi^\dg(\vr'))_j] = \dl_{ij}\dl^{(3)}\!(\vr - \vr')
 = \la\vr|\dl_{ij}|\vr'\ra,\enskip i,j = 1,2,3,
}{10a}
which is the straightforward consequence of Hermitian quantization of
the real \e{canonical} vector fields $\vPhi(\vr)$ and $\vPi(\vr)$ that
accords with the Dirac relation of commutators to classical Poisson
brackets for the quantized components of the classical phase-space
vector.  Here, however, we must contend with the technical/mathemati%
cal annoyance imposed by the \e{transverse}-vector character of the
real canonical fields $\vPhi(\vr)$ and $\vPi(\vr)$, which effectively
\e{removes} one third of the classical phase-space degrees of freedom
that would have been present had $\vPhi(\vr)$ and $\vPi(\vr)$ been
real canonical \e{full}-vector fields.  What is \e{missing} from the
classical \e{full}-vector field phase space is, of course, its
\e{longitudinal part}, i.e., all those vector fields which can be
written as the gradient of a scalar field.  The $ij$ \e{components}
of the \e{projection operator} into this \e{longitudinal part} of the
space of vector fields are given by the integro-differential operators
$(-\pa_i(-\del^2)^{-1}\pa_j)$.  This operator is readily verified to
map any vector field into the gradient of a scalar, to be a symmetric
linear operator on the natural Hilbert space of real vector fields and
to be equal to the square of itself.  Therefore the $ij$ components of
the projection operator into the \e{transverse part} of the space of
vector fields are given by the operators $(\dl_{ij} +
\pa_i(-\del^2)^{-1}\pa_j)$.  Thus to be mathematically consistent with
the \e{transverse}-vector nature of the classical canonical and quantum
operator fields, we need to replace Eq.~(10a) by the commutation rela%
tion,
\re{
[(\qvPsi(\vr))_i, (\qvPsi^\dg(\vr'))_j] =
\la\vr|(\dl_{ij} + \pa_i(-\del^2)^{-1}\pa_j)|\vr'\ra = \\
\null \\
\dl_{ij}\dl^{(3)}\!(\vr - \vr') - (2\pi)^{-3}\int e^{i\vk\dt(\vr - \vr')}
(\vk)_i|\vk|^{-2}(\vk)_j\,d^3\vk\,,\enskip i,j = 1,2,3.
}{10b}
This commutation relation has the well-known interpretation that
$\qvPsi^\dg(\vr)$ creates a free-photon state localized at $\vr$,
while $\qvPsi(\vr)$ annihilates such a free-photon state, so that the
underlying Hilbert space, called Fock space, accommodates arbitrarily
large numbers of free photons~[6].  The real-valued bilinear Hamil%
tonian functional $H[\vPsi^\ast, \vPsi]$ of Eq.~(9c) must be corres%
pondingly quantized to become the Hermitian Hamiltonian \e{operator}
$\qH[\qvPsi^\dg, \qvPsi]$ that governs the time evolution of this
second-quantized photodynamical system of arbitrarily-large numbers
of free-photons.  An very important (and normally expected) conse%
quence of this is that in the Heisenberg picture the non-Hermitian
field $\qvPsi$ \e{obeys the selfsame Schr\"{o}dinger equation} of Eq.%
~(9b) \e{that was its equation of motion before it was second-quan%
tized}.  In other words, the occurrence \e{at the quantized-field lev%
el} of the free-photon Schr\"{o}dinger equation of Eq.~(9b) is \e{an
unavoidable consequence of correct quantization of the source-free
Maxwell equations}.

At this second-quantized level, $\vA_T$, the transverse part of the
electromagnetic potential, \e{also} becomes an operator, in fact a
Hermitian one.  From Eq.~(9d) we see that its operator form will be
given by,
\re{
\qvA_T =(\h c/2)^\hf(-\del^2)^{-\qr}\,(\qvPsi + \qvPsi^\dg).
}{10c}
It is apparent that $\qvA_T$ is the object which couples photons to
charged particles.  From its form as given by Eq.~(10c) it is clear
that charged particles can both absorb and emit photons via $\qvA_T$.
We now cast a backward glance at what has been accomplished in the
foregoing sections, and then take to its completion the discussion
we have just begun on how to theoretically set up the electromagnetic
interactions that affect charged particles.

\sct{Photodynamical and coulombic interactions of charged particles}
\noindent
The source-free case of electromagnetism has, after its meticulous
canonical Hamiltonization, which is an absolute necessity that has
been essentially universally honored only in the breach for the past
century and a half, effortlessly yielded up the completely natural
first- and second-quantized theories of free transverse-vector pho%
tons, with the later governed by the bilinear-field Hamiltonian opera%
tor $\qH[\qvPsi^\dg, \qvPsi]$, which is the straightforward quantized
version of the c-number Hamiltonian functional $H[\vPsi^\ast, \vPsi]$
of Eq.~(9c).  The quantized-field Hamiltonian $\qH[\qvPsi^\dg,\qvPsi]$
yields in the Heisenberg picture an equation of motion for the quan%
tized field $\qvPsi$ that is \e{form-identical} to Eq.~(9b), the
Schr\"odinger equation for the first-quantized photon wave function
$\vPsi$.  This \e{second-quantized} form of the Schr\"{o}dinger equa%
tion for $\qvPsi$ obviously \e{still features the very same} $m\rta 0$
limit of the relativistic \e{square-root} Hamiltonian operator $(m^2
c^4 + |c\qvp|^2)^\hf$ (which comes to $\h c(-\del^2)^\hf$ in configu%
ration representation) that it featured as it applied to $\vPsi$.
Thus the emergence of this free-photon Schr\"{o}dinger equation with
the just-mentioned \e{natural square-root relativistic Hamiltonian op%
erator} is the \e{unavoidable consequence} of quantization of source-%
free electromagnetism!

Maxwell's equations indicate that charged matter interacts with the
dynamical transverse part of electromagnetism via a global transverse
current density, as is seen from Eq.~(3c).  Microscopic theories of
charged particles, however, indicate that their interaction with
transverse electromagnetism occurs via $\vA_T$, the transverse part of
the electromagnetic potential, whose quantized form $\qvA_T$ is given
directly in terms of the Hermitian part of the non-Hermitian quantized
photon wave-function field $\qvPsi$ by Eq.~(10c).

To describe charged particles and electromagnetism participating in
mutual interaction, we require a Hamilton which consists of the
\e{sum} of $\qH[\qvPsi^\dg, \qvPsi]$ with a Hamiltonian that describes
the charged particles, \e{specifically including} the effects on them
which an \e{arbitrary external c-number four-vector potential} $A^\mu
= (A^0, \vA)$ produces.  After writing $\vA = \vA_L + \vA_T$ by apply%
ing the now familiar unique linear longitudinal/traverse decomposition
of any vector field, we \e{concretely replace} $\vA_T$ by the Hermi%
tian photon field operator $\qvA_T$ that is given by Eq.~(10c).  The
\e{remaining} $A^0$ and $\vA_L$ \e{have no direct relation to pho%
tons}: in light of the basic definition $\vE_L = (-\del A^0 -
\dot\vA_L/c)$ and Eq.~(3a), they are determined 1) by the global
charge density $\rho$ of the system, \e{which exerts a coulombic ef%
fect on itself}, and 2) by one's choice of gauge.  The system's global
charge density $\rho$ is given by the functional derivative of the
charged-particle Hamiltonian with respect to $A^0$. If it should hap%
pen that this $\rho$ \e{itself} depends on $A^0$, then, after making
the choice of gauge, one will \e{still not have in hand the actual re%
sult for} $A^0$, but only an \e{implicit equation} for that result,
whose solution can probably only be successively approximated by iter%
ation.  So far as choice of gauge is concerned, by far the simplest is
the Coulomb gauge, which, via its requirement that $\del\dt\vA = 0$,
neatly implies that $\vA_L$ vanishes, albeit this is \e{prima facie}
disrespectful of special relativistic precepts.  That notwithstanding,
there does not seem to be any physical reason to eschew the Coulomb
gauge, since Maxwell's equations yield the result of the coulombic ef%
fect to be the Eq.~(3a) form of $\vE_L$ in terms of $\rho$, which is
\e{very closely related} to the Coulomb gauge result below for $A^0$,
when $A^0$ is \e{additionally} required to be linear and \e{homogen%
eous} in $\rho$, namely,
\re{
A^0(\vr, t) = \hf\int (4\pi|\vr - \vr'|)^{-1}\rho(\vr', t)\,d^3\vr'.
}{11}
An \e{unusual factor of one half} has been inserted into Eq.~(11) to
\e{compensate the double counting} that would otherwise occur because
this particular $A^0$ inherently interacts coulombically \e{with the
very same charge density} $\rho$ \e{that gives rise to it}.

If, in spite of the very close relationship of the Coulomb gauge re%
sult of Eq.~(11) to the Maxwell equation result of Eq.~(3a) for $\vE_L
$, a relativistically more plausible gauge is nonetheless thought to
be desirable, the \e{retarded Lorentz gauge} of Ref.~[3] would seem to
be an excellent choice.  Scrapping the Coulomb gauge requirement that
$\vA_L = \vz$, but imposing the relativistically impeccable Lorentz
condition $(c\del\dt\vA_L + \dot A^0) = 0$ implies that $A^0$ satis%
fies the second-order in time partial differential equation $(\ddot
A^0/c^2 - \del^2A^0) = \rho$.  Imposing the \e{further} requirements
that $A^0$ be linear and \e{homogeneous} in $\rho$, and that it res%
pond to \e{changes} in $\rho$ \e{only after} the $c$-speed \e{retarda%
tion time} $|\vr - \vr'|/c$, turns out to yield an $A^0$ which is
\e{different} from that of Eq.~{11} \e{only in that} $\rho(\vr',t)$ on
its right-hand side \e{is replaced by} $\rho(\vr', t - |\vr -\vr'|/c)
$.  With that re\-tard\-ed-Lo\-rentz-gauge result for $A^0$ in terms
of $\rho$ \e{in hand}, one then determines the re\-tard\-ed-Lo\-rentz%
-gauge $\vA_L$ from the re\-tard\-ed-Lo\-rentz-gauge $\dot A^0$ and
the Lo\-rentz condition, which explicitly yields $\vA_L = \del[(-\del%
^2)^{-1}(\dot A^0/c)]$.  For the case that $\rho$ is \e{time-in\-de\-%
pen\-dent}, this full re\-tard\-ed-Lo\-rentz-gauge result \e{reduces
to that of the Coulomb gauge}.

\sct{Conclusion}
\noindent
It has recently been strenuously argued on the basis of the corres%
pondence principle that the \e{only physically sensible} Hamiltonian
operator for a solitary, relativistic first-quantized free particle
of positive mass $m$ is the square root operator $(m^2c^4 + |c\qvp|^2
)^\hf$~[7].  The extension of this idea to massless particles of
course assigns them the Hamiltonian operator $|c\qvp|$, which is
$\h c(-\del^2)^\hf$ in configuration representation.  One naturally
turns to a long and firmly established theory, namely source-free
electromagnetism,  which is supposed to include a massless free par%
ticle, namely the free photon, in its ambit, to see how this Hamilton%
ian-operator cum Schr\"{o}dinger equation idea fares in the context of
its quantization.  The foregoing work shows that it fares absolutely
brilliantly, with the expected Schr\"{o}dinger equation and its asso%
ciated Hamiltonian operator being perfectly maintained right to the
level of the quantized field (or, more precisely, quantized photon
wave function) \e{if} one \e{does not neglect} to properly canonically
Hamiltonize the source-fee Maxwell equations (whose electric and mag%
netic fields are \e{very far} from being properly canonical!) \e{be%
fore} embarking on quantization.  In fact, this Schr\"{o}dinger equa%
tion's appearance at the \e{first-quantized} level turns out to be
\e{no more} than a simple, direct \e{by-product} of merely the \e{pro%
per canonical Hamiltonization} of Maxwell's source-free equations!

These results for source-free electromagnetism lend impressive support
to the almost ridiculously straightforward idea that the \e{correspon%
dence-principle-mandated} relativistic free-particle square-root Ham%
iltonian operator $(m^2c^4 + |c\qvp|^2)^\hf$ and associated time-de%
pendent Schr\"{o}dinger equation is \e{without exception} the correct
\e{starting point} for relativistic quantum theory (a glance at the
section just above, and perhaps even more so at the latter sections of
Ref.~[7], reveals a taste of the mind-boggling complexity and richness
which this \e{mere starting point} rapidly gives way to).  Who would
be prepared to \e{for an instant} contest the completely parallel as%
sertion that the correspondence-principle-mandated nonrelativistic
free-particle kinetic-energy Hamiltonian operator $|\qvp|^2/(2m)$ and
associated time-dependent Schr\"{o}dinger equation is \e{without ex%
ception} the correct \e{starting point} for nonrelativistic quantum
theory?  It may not, in this very regard, have escaped the reader's
attention that the correspondence-principle-mandated \e{square root}
relativistic Hamiltonian has the \e{delicately subtle} property that
as $c\rta\infty$,
\de{
[(m^2c^4 + |c\qvp|^2)^\hf - mc^2]\,\rta\,|\qvp|^2/(2m).
}
For the linearized Dirac Hamiltonian $\vec\al\dt\qvp c + \bt mc^2$,
there is simply \e{no remotely similar} property.  The \e{very best}
in this regard which can be done with the Dirac Hamiltonian is to sub%
tract away its value at $\qvp = \vz$, leaving $\vec\al\dt\qvp c$,
which \e{diverges} as $c\rta\infty$!  Klein-Gordon theory fails alto%
gether to present us with a Hamiltonian (which is the proximate cause
of its quantum-theoretic downfall~[7]), but it does give us the
\e{square} of the correspondence-principle-mandated relativistic one,
namely $(m^2c^4 + |c\qvp|^2)$.  Subtracting away its value at $\qvp =
\vz$ leaves $|c\qvp|^2$, which \e{also diverges} as $c\rta\infty$!
The relativistic free-particle \e{square root Hamiltonian} $(m^2c^4 +
|c\qvp|^2)^\hf$ is \e{very subtly tuned} indeed!  The square root
\e{itself} is in fact theoretically \e{completely entwined} with the
square roots that are \e{archetypal} of the Lorentz transforma%
tion~[7].

In contrast to this square root, which is \e{uniquely fathered} for
the free particle by the Lorentz transformation \e{itself}~[7], both
the Klein-Gordon and Dirac equations were \e{artificially concocted}
for the \e{express purpose} of evading nonlocal integral operators in
the configuration representation of relativistic quantum mechanics~[8,
7].  In light of this fact, it is a truly monumental irony that while
the source-free Maxwell equations written in terms of the electric and
magnetic fields achieve \e{exactly the configuration-representation
locality} so willfully \e{insisted upon} by Klein, Gordon and Dirac,
those equations are thereby \e{inconsistent with quantization}, being
\e{far from properly canonically Hamiltonized}, and when this issue is
duly attended to, they \e{utterly lose} that ``precious'' configura%
tion-representation locality to \e{precisely} one of the square-root
operators which Klein, Gordon and Dirac so abjured!  An extremely pow%
erful lesson emerges from this: there quite simply can be no physical%
ly legitimate union of special relativity with quantum theory \e{with%
out} those square-root operators.  The efforts of Klein, Gordon and
Dirac to rid relativistic quantum theory of these square-root opera%
tors and achieve configuration-representation locality merely engen%
ders a dismal list of theoretically inappropriate or grotesquely un%
physical consequences~[7].  The Dirac free particle theory, for exam%
ple, presents a number of operators that are \e{without question}
physical observables, such as the three components of velocity and the
energy (i.e., the Dirac Hamiltonian itself), which nonetheless
\e{fail} to mutually commute when the limit $\h\rta 0$ is taken!  That
is grotesquely unphysical, and quite enough to permanently consign Di%
rac theory to the dustbin.  But just for good measure, these commuta%
tors \e{diverge} in the nonrelativistic limit $c\rta\infty$!  Bereft
of a Hamiltonian, Klein-Gordon theory has no time evolution operator
and no Heisenberg picture.  For the same underlying reason, it mani%
fests negative probabilities.  In short, it is so hopelessly crippled
that it cannot be regarded as quantum theory at all.  These are items
from the dismal list of Dirac and Klein-Gordon equation shortcomings,
but there is not so much as a \e{single item} in that pejorative list
which pertains to the \e{square root Hamiltonian}!  With this \e{ut%
terly lopsided} accounting of the theoretical pros and cons, it is far
past time for the theoretical physics community to finally awaken to
just what in its repertoire needs to be revised~[7] to give the opera%
tor $(m^2c^4+|c\qvp|^2)^\hf$ \e{exactly the same standing} in relati%
vistic quantum physics that the operator $|\qvp|^2/(2m)$ properly has
in nonrelativistic quantum physics.

\vskip 1.75\baselineskip\noindent{\frtbf References}
\vskip 0.25\baselineskip

{\parindent = 15pt
\sk\item{[1]}
S. K. Kauffmann,
arXiv:0910.2490 [physics.gen-ph]
(2009).
\sk\item{[2]}
S. K. Kauffmann,
arXiv:0908.3755 [quant-ph]
(2009).
\sk\item{[3]}
S. K. Kauffmann,
arXiv:1005.1101 [physics.gen-ph]
(2010).
\sk\item{[4]}
L. I. Schiff,
\e{Quantum Mechanics}
(McGraw-Hill, New York, 1955).
\sk\item{[5]}
J. D. Bjorken and S. D. Drell,
\e{Relativistic Quantum Fields}
(McGraw-Hill, New York, 1965).
\sk\item{[6]}
S. S. Schweber,
\e{An Introduction to Relativistic Quantum Field Theory}
(Harper \& Row, New York, 1961).
\sk\item{[7]}
S. K. Kauffmann,
arXiv:1009.3584 [physics.gen-ph]
(2010).
\sk\item{[8]}
J. D. Bjorken and S. D. Drell,
\e{Relativistic Quantum Mechanics}
(McGraw-Hill, New York, 1964).
}
\bye